# Engineering spin-orbit torque in Co/Pt multilayers with perpendicular magnetic anisotropy

Kuo-Feng Huang,[1] Ding-Shuo Wang,[1] Hsiu-Hau Lin,[2] and Chih-Huang Lai[1]*

[1]Department of Materials Science and Engineering, National Tsing Hua University, Hsinchu 30013, Taiwan
[2]Department of Physics, National Tsing Hua University, Hsinchu 30013, Taiwan

**To address thermal stability issues for spintronic devices with a reduced size, we investigate spin-orbit torque in Co/Pt multilayers with strong perpendicular magnetic anisotropy. Note that the spin-orbit torque arises from the global imbalance of the spin currents from the top and bottom interfaces for each Co layer. By inserting Ta or Cu layers to strengthen the top-down asymmetry, the spin-orbit torque efficiency can be greatly modified without compromised perpendicular magnetic anisotropy. Above all, the efficiency builds up as the number of layers increases, realizing robust thermal stability and high spin-orbit-torque efficiency simultaneously in the multilayers structure.**

* Correspondence and requests for materials should be addressed to C.-H. L. (chlai@mx.nthu.edu.tw)

Manipulating magnetization by electrical means, such as spin-transfer torque[1-3] or electric-field control of magnetism[4-6], is an important task for realizing spintronic devices. Recently, the so-called spin-orbit-torque (SOT) has attracted intense attention because it provides an efficient way for manipulating magnetization and offers a new route for spintronic device designs. The spin-orbit torque arises mainly from two distinct origins: the spin current generated by spin Hall effect,[7] or the effective magnetic field induced by Rashba interaction.[8] By simply applying a charge current through the asymmetrical structure composed of a ferromagnet (FM) layer and adjacent heavy metal layers with strong spin-orbit coupling, spin-orbit torque can be generated in the FM and drives the magnetization reversal.[9] In consequence, SOT has been proposed as an alternative writing mechanism for the next generation magnetoresistive random access memory (MRAM),[7, 10] especially to manipulate the FM layer with perpendicular magnetic anisotropy (PMA) for better thermal stability.[11-13] Nevertheless, most of the existing works are restricted to the single FM layer around 1 nm in thickness for preserving the PMA, where the relatively poor thermal stability due to the limited volume of FM layer is a big concern when scaling down to several tens of nanometers. In this work, we would like to evaluate the SOT in the Co/Pt multilayers (MLs), which possesses good enough PMA with tunable volume of the FM layer and also shows a good magnetoresistance in MgO-based magnetic tunnel junctions with the interfacial CoFeB insertion.[14, 15] Although a high SOT has been reported on the Co/Pd MLs recently,[13] no systematic works have been done for explaining the presence of SOT in the multilayers, where the spin currents from heavy metal may counteract each other and no Rashba interaction can be seen without structure inversion asymmetry. By conducting the lock-in technique,[16-18] the dependence of SOT efficiencies of the Co/Pt MLs on different repeating layer number (N) is investigated. Besides, Cu and Ta are inserted into the Co/Pt MLs for modifying the interfacial conditions to clarify the origin of SOT in multilayers system and to tune the strength of SOT efficiency. We found that the magnitude of SOT in Co/Pt MLs strongly depends on the interfacial conditions. In addition, as the numbers of repeating layers increases, the FM volume and SOT efficiency increase accordingly, showing great potential for improving the thermal stability and the writability.

By DC magnetron sputtering, we first prepared the Co/Pt MLs, sub./Ta(2.5 nm)/[Pt(2 nm)/Co(0.9 nm)]$_N$/Pt(2 nm)/Ta(2.5 nm), with varied repeating layer numbers (N= 1, 2, 4, 8) and the Pt/Co/Pt tri-layers with varied Pt thicknesses, sub./Ta(2.5 nm)/Pt(x nm)/ Co(0.9 nm)/Pt(x nm)/Ta(2.5 nm), on thermally oxidized Si (100) substrates. Furthermore, 0.5 nm Cu or 0.15 nm Ta was inserted into some of the Co/Pt MLs on the top of each Co layers, denoted as Co/Cu/Pt and Co/Ta/Pt MLs, respectively. By a vibrating sample magnetometer (VSM), we confirmed that all of those film structures show perpendicular magnetic anisotropy and extracted both the saturation magnetization-thickness product ($M_S*t$, acquired by normalizing the total magnetization by its area) and the magnetic anisotropy field ($H_K$) for calculating the corresponding magnetic anisotropy energy ($K_{eff} = M_S H_K/2$) from the in-plane and out-of-plane hysteresis loops. Then, these multilayers were patterned into Hall-bar (5μm in width) by traditional photolithography and reactive-ion etching. On all of the Hall-bars, we further fabricated the Ta(10 nm)/Pt(100 nm) electrodes by photolithography, DC magnetron sputtering, and lift-off process sequentially to conduct the following measurements. To study the SOT quantitatively, the lock-in measurements with planar Hall effect correction[18] were conducted on the Hall-bar structure for acquiring the in-phase first harmonic ($V_\omega$) and the out-of-phase second harmonic ($V_{2\omega}$) Hall voltages. We defined the longitudinal direction as the direction of the applied current and transverse direction as the directions transverse to the current, as shown in Figure 1(a). The effective field along longitudinal and transverse directions ($\Delta H_L$ and $\Delta H_T$, respectively) can therefore be quantitatively evaluated from the measured $V_\omega$ and $V_{2\omega}$ based on the formula derived by Hayashi *et al.*[18]

We first evaluated the SOT in the [Co (0.9 nm)/Pt (2 nm)]$_8$ MLs by adopting the lock-in measurement[18] to measure the magnitude and sign of the longitudinal ($\Delta H_L$) and transverse ($\Delta H_T$) effective field with varied charge current density ($J_e$). The measurement setup is shown in Figure 1(a) and the $J_e$ dependence of $\Delta H_{L(T)}$ is shown in Figure 1(b). The magnitude of both $\Delta H_L$ and $\Delta H_T$ increase linearly with $J_e$, where the longitudinal and transverse spin-orbit-torque efficiencies ($\beta_{L(T)}$), defined as $\Delta H_{L(T)}/J_e$, can therefore be extracted from the slope. The $\beta_L$ and $\beta_T$ of the [Co/Pt]$_8$ sample ($K_{eff}$ ~ 4.6x10$^6$ erg/cm$^3$, $M_s$*t ~ 6.7x10$^{-4}$ emu/cm$^2$) are 62 and 24 Oe-cm$^2$/10$^7$A, respectively, showing a comparable $\beta_{L(T)}$ and $K_{eff}$ to the cases with single FM layer,[11, 19] but with largely enhanced $M_s$*t, that is, improved thermal stability.[15] Besides, because the $\beta_L$ is 2.6 times larger than the $\beta_T$, the damping-like torque arisen by the spin current from the giant spin Hall effect of Pt layers should be the main source of the SOT in the Co/Pt MLs.[17, 20] Note that the symmetric structure of the Co/Pt MLs implies null SOT due to the cancellation of the spin currents from different Pt layers. To address the unexpected SOT observed in Co/Pt MLs, we focus on the longitudinal SOT efficiency $\beta_L$ first.

To simplify the system, we started with a series of tri-layers, Pt (x nm)/Co (0.9 nm)/Pt (x nm), the basic repeating units comprising Co/Pt MLs, where the thicknesses of the top and bottom Pt are equal, denoted as a symmetric tri-layer. Due to the equal thickness of the top and bottom Pt, the bulk contribution of the spin current from the Pt layers due to spin Hall effect is expected to be the same magnitude but with opposite spin directions, cancelling each other with no $\beta_L$.[11, 21] However, as shown in Figure 1(c), we still observed significant $\beta_L$ (about 21 ~ 33 Oe-cm$^2$/10$^7$A) in the symmetric tri-layers, indicating the spin currents from the Pt layers are not completely cancelled. We speculate that the difference between top and bottom interfaces is the key for the presence of $\beta_L$ in the symmetric tri-layers. The different interfacial structures can enormously change the transport of spin current, for example, modifying the ratio of spin backflow[22, 23] or changing the degree of spin-flip scattering.[24] Considering the interfaces of Co/Pt MLs, the different interfacial structures between the Pt/Co and the Co/Pt were reported in the sputtered Co/Pt MLs;[25] the bottom Pt/Co shows chemically sharp interfaces but the top Co/Pt suffers from the intermixing during sputtering, which may be the reason for the presence of $\beta_L$. Furthermore, as shown in Figure 1(d), with increasing top Pt thickness (bottom keeps 2nm), denoted as asymmetric tri-layers, we found a monotonic increase of the magnitude of $\beta_L$ without sign change. The $\beta_L$ increase can be attributed to the enhancement of spin current density from the thicker top Pt layer, leading to an effective net spin current. Therefore, since the sign of $\beta_L$ are all identical for the symmetric and asymmetric tri-layers, we think the effective net spin current in the symmetric tri-layers, which leads to the unexpected $\beta_L$, is also from the top, suggesting a higher transmission ratio of spin current through the top Co/Pt interface. The higher transmission ratio of spin current for the top Co/Pt interface can also be found in the tri-layers.[26]

To verify the $\beta_L$ modification due to the discrepancy between top and bottom interfaces in multilayers, we prepared a series of Co/Pt MLs with different repeating layer numbers (N= 1, 2, 4, and 8), and inserted Cu or Ta in some of the Co/Pt MLs at the top interfaces of each Co in the multilayers for altering the properties of top interfaces but keeping the bottom intact. For these multilayers, we found that all the $M_S*t$ increases linearly with N, as shown in Figure 2(a). The $M_S*t$ reveals slight reduction when the Cu or Ta was inserted, implying the top interfaces are modified by the insertions. On the other hand, although the $K_{eff}$ shows a slight increase with increasing N or with the extra Cu insertion, the $K_{eff}$ remains in the same level with good PMA, as shown in Figure 2(b), indicating that the bottom interfaces (the origin of the PMA)[27] are preserved. Figures 3(a)~3(c) show the measured $\beta_L$ and $\beta_T$ of Co/Pt, Co/Cu/Pt, and Co/Ta/Pt MLs with increasing N respectively. We found that the $\beta_L$ was enhanced with the extra Ta insertions, but degraded with the Cu insertions. Besides, the sign of $\beta_L$ remains the same for all samples, indicating the effective net spin current generated by the top Pt layers dominates. Note that the insertion of Cu at Co/Pt interfaces has been reported to degrade the transmission of spin current through the interface either due to the enhanced spin-flip scattering[24] or the suppressed spin mixing conductance.[23, 28, 29] As a result, the decreased $\beta_L$ may be attributed to the reduction of spin current transmission through the top interfaces, which lowers the disparity between top and bottom interfaces. On the contrary, regarding the Ta insertions, possessing a giant spin Hall angle ($\theta_{SH}$ ~ -0.12)[7] but with the opposite sign to Pt, the combination of Ta and Pt are expected to counteract each other and to reduce the $\beta_L$, which is contradictory to our observations: $\beta_L$ increased from 62 Oe-cm$^2$/10$^7$A for the [Co/Pt]$_8$ MLs to 88 Oe-cm$^2$/10$^7$A for the [Co/Ta/Pt]$_8$ MLs, shown in Figure 3(c). We suggest that the inserted thin Ta layers react with the Co, forming a magnetic dead layer, as observed in Figure 2(a), where the $M_S$*t of the Ta inserted samples show 10% drop compared to the corresponding Co/Pt MLs with the same N. The randomness in the magnetic dead layer may alter the spin-mixing conductance at the interfaces[22, 30] and leads to a more transparent top interface with larger discrepancy between the top and bottom interfaces based on the increased $\beta_L$. Regarding the case in pure Co/Pt MLs, we speculate that the intermixing of the top Co/Pt interfaces may also alter the spin-mixing conductance as the magnetic dead layer introduced by Ta insertion layers does, which gains the transparency of spin current, leading to a discrepancy between the chemical sharp bottom and the top interface with severe intermixing. Consequently, we believe that the $\beta_L$ are sensitive to the discrepancy between top and bottom interfaces in the multilayers system and can be tuned by modifying the interfaces.

In addition, we also found that the $\beta_L$ enhanced with increasing N in all MLs with different insertion materials, as shown also in Figures 3(a)~3(c). The enhanced $\beta_L$ indicates that the effective net spin current absorbed by each Co layer increases with N, where the total Pt thickness also increases even though the Pt layers are separated by the 0.9 nm Co layers. That is, the adjacent Pt layers may couple to each other across the Co layer to enhance the driving force of the spin current. To understand this N dependence of $\beta_L$ in Co/Pt MLs, we need to consider the spin diffusion length of Co ($l_{sf,Co}$) and Pt ($l_{sf,Pt}$) first, where Co possesses a relatively long $l_{sf,Co}$ ( > 30 nm)[31] compared to the Pt (~ 4 nm). In the Co/Pt MLs, the spin accumulation at the interfaces due to spin Hall effect is the main driving force for the spin current flowing into each Co layers. When the FM layer ($d_F$) is thick enough ($d_F \gg l_{sf,Co}$), where the spin accumulation at the interfaces can be totally relaxed or the spin current can be totally dissipated in the FM layer, the spin accumulations of the adjacent Pt layers are independent to each other; therefore, the spin accumulation and spin current may depend only on the thickness of each Pt layer and the interface conditions mentioned above. However, when the FM layer is thin ($d_F < l_{sf,Co}$), the spin accumulation cannot be totally relaxed within such a thin FM layers and the spin current from one side may affect the one from the other side, leading to the coupling of the spin accumulation between Pt layers. The coupling of the Pt layers across Co layers enhances the spin accumulation at the interfaces, effective to the increase of the Pt thickness, and thus leads to the $\beta_L$ enhancement.

Regarding the relatively small $\beta_T$, it can be explained by the lack of structure inversion asymmetry in the all metallic Co/Pt structure as well as the relatively thick Co layers (0.9 nm) compared to the Co (0.2 nm)/Pd case,[13] which may hinder the Rashba interaction in the Co/Pt MLs. Besides, we found that $\beta_T$ of all samples scales with its corresponding $\beta_L$, which is induced by spin Hall effect. That is, the varied $\beta_T$ of different Co/Pt MLs or tri-layers might also be originated from spin Hall effect, which is also found in other material systems,[12, 17] where the reflection of the spin current may contribute to the $\beta_T$.[20]

In summary, in Co/Pt MLs, because of the combination of the spin accumulation of the coupled Pt layers across the Co layer and the discrepancy between top and bottom interfaces, the $\beta_L$ increases with increasing N and can be tuned by modifying its interfaces without sacrificing the perpendicular magnetic anisotropy. The Ta insertion on the top of each Co layer enhances the discrepancy between top and bottom interfaces, leading to larger $\beta_L$ compared to the pure Co/Pt MLs. On the other hand, Cu insertion reduces the interface discrepancy with smaller $\beta_L$. Consequently, we believe that the multilayers setup is of great potential for real applications because the enhancing SOT efficiency can be achieved without compromise of thermal stability.

This work was partially supported by the Ministry of Science and Technology of Republic of China under Grant No. NSC 102-2221-E-007-043-MY2 and MOST 104-2221-E-007-047-MY2.

**Figure captions**

**Figure 1.** (a) Sketch of the lock-in measurement setup for measuring the spin-orbit-torque effective fields in both of the longitudinal ($\Delta H_L$) and the transverse ($\Delta H_T$) directions. (b) $\Delta H_L$ and $\Delta H_T$ of the Ta(2.5nm)/[Co(0.9nm)/Pt(2)]$_8$/Ta(2.5nm) multilayers after planar Hall effect correction with different charge current density ($J_e$). The longitudinal ($\beta_L$) and transverse ($\beta_T$) spin-orbit-torque efficiencies, defined as $\Delta H_{L(T)}/J_e$ of (c) the symmetric tri-layers film structures and (d) the tri-layers with thicker varied top Pt layer, with positive/negative initial magnetization states (+/-M).

**Figure 2.** Variations with increasing repeating layer numbers, N, of (a) saturation magnetization-thickness products ($M_S*t$) and (b) effective magnetic anisotropy energy ($K_{eff}$) for [Pt(2 nm)/ Co (0.9 nm)]$_N$, [Pt(2 nm)/ Cu (0.5nm)/ Co (0.9 nm)]$_N$, and [Pt(2 nm)/ Ta (0.15nm)/ Co (0.9 nm)]$_N$ MLs.

**Figure 3.** Longitudinal ($\beta_L$) and transverse ($\beta_T$) spin-orbit-torque efficiencies of (a) [Co/Pt]$_N$ multilayers, (b) [Co/Cu/Pt]$_N$ multilayers, and (c) [Co/Ta/Pt]$_N$ multilayers with different repeating layer numbers, N, and positive/negative initial magnetization states (+/-M).

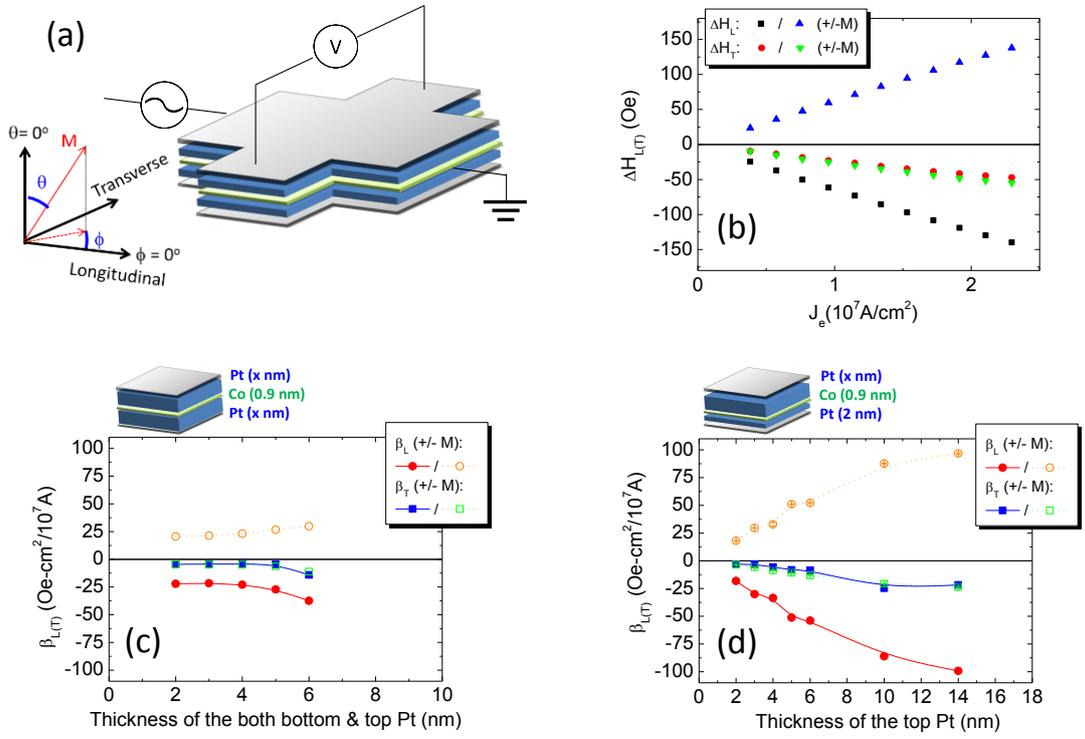

**Figure 1.** Kuo-Feng Huang

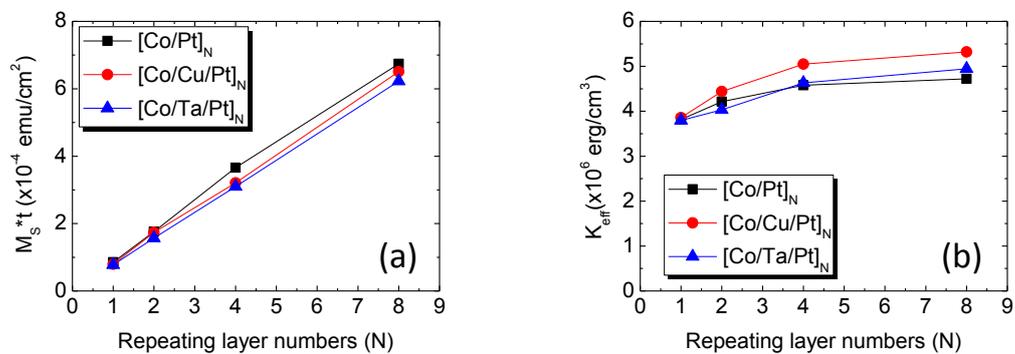

**Figure 2.** Kuo-Feng Huang

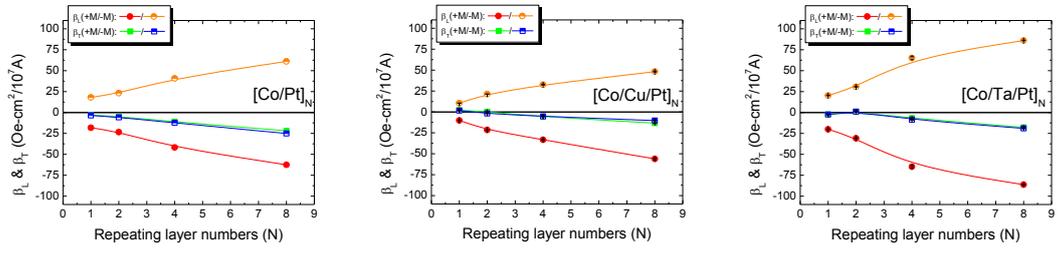

**Figure 3.** Kuo-Feng Huang